%% file: paper.tex
\documentclass[letterpaper,twocolumn,10pt]{article}
\usepackage{usenix-2020-09}
\usepackage[toc,page]{appendix}

\usepackage{tikz}
\usepackage{amsmath}
\usepackage{multirow}
\usepackage{graphicx}
\usepackage{bbm}
\usepackage{xurl}

\usepackage{filecontents}
\usepackage{authblk}

\newcommand{\name}{SecEncoder}

\begin{document}

\date{}

\title{\Large \emph{\name{}:} \bf Logs are All You Need in Security}


\author{Muhammed Fatih Bulut}
\author{Yingqi Liu}
\author{Naveed Ahmad}
\author{Maximilian Turner}
\author{Sami Ait Ouahmane}
\author{Cameron Andrews}
\author{Lloyd Greenwald}
\affil{Microsoft Security AI Research \authorcr \small{\tt \{mbulut, yingqiliu, navahm, maxturner, saitouahmane, camandrews, lgreenwald\}@microsoft.com}}

\maketitle

\input{abstract.tex}
\input{intro.tex}
\input{related.tex}

\input{system.tex}
\input{evaluation.tex}

\input{use_cases.tex}

\input{discussion.tex}

\input{conclusion.tex}

\section*{Acknowledgments}

We would like to acknowledge the following individuals for their valuable contributions to the development and evaluation of \name{}: Aditi Shah, Anush Sankaran, Brent Miller, Bryan Xia, Bryan Jeffrey, Evan Greenberg, Hani Neuvirth-Telem, Hany Gerges, Leo Betthauser, Likhita Manchikanti, Malachi Jones, Matthew Duncan, Noa Nutkevitch, Sam Yang, Shani Klein Antman, Siyue Wang, Tanvi Raja, Yossi Bruss, Vignesh Nayak and many more.

\small
\bibliographystyle{plain}
\bibliography{paper.bib}

\input{appendices.tex}

\end{document}

%% file: abstract.tex
\begin{abstract}
Large and Small Language Models (LMs) are typically pretrained using extensive volumes of text, which are sourced from publicly accessible platforms such as Wikipedia, Book Corpus, or through web scraping. These models, due to their exposure to a wide range of language data, exhibit impressive generalization capabilities and can perform a multitude of tasks simultaneously. However, they often fall short when it comes to domain-specific tasks due to their broad training data. This paper introduces \name{}, a specialized small language model that is pretrained using security logs. \name{} is designed to address the domain-specific limitations of general LMs by focusing on the unique language and patterns found in security logs. 
Experimental results indicate that \name{} outperforms other LMs, such as BERT-large, DeBERTa-v3-large and OpenAI's Embedding (text-embedding-ada-002) models, which are pretrained mainly on natural language, across various tasks. Furthermore, although \name{} is primarily pretrained on log data, it outperforms models pretrained on natural language for a range of tasks beyond log analysis, such as incident prioritization and threat intelligence document retrieval. This suggests that domain-specific pretraining with logs can significantly enhance the performance of LMs in security. These findings pave the way for future research into security-specific LMs and their potential applications. 
\end{abstract}

%% file: intro.tex
\section{Introduction}
\label{sec:intro}

Transformers \cite{vaswani2017attention} are a breakthrough AI architecture that have facilitated the development of various Language Models (LMs) \cite{DBLP:journals/corr/abs-2108-07258, ibmFoundationModels}. These models can harness huge amounts of data to perform diverse tasks across language and other modalities such as audio, image and video. Some examples of LMs are BERT, RoBERTa, GPT, Gemini and PaLM \cite{liu2019roberta, chowdhery2022palm, devlin2019bert, brown2020language, geminiteam2023gemini, 2024gemini15, openai2023gpt4, gpt_4o}. LMs can differ in their size, data sources, learning objectives, and are trained on large collections of text, such as Wikipedia, books, news articles, code, social media posts or web scraping. Certain language models, such as encoder-only models, are designed to encode the semantic and syntactic information of natural language into high-dimensional vectors, enabling them to perform downstream tasks like search, classification, summarization, translation, and question answering. Decoder-only models, on the other hand, are capable of generating natural language text by sampling from their probability distributions, producing coherent and fluent outputs. Some models, such as GPT-o and Gemini 1.5, extend these capabilities to multimodal content generation, creating images, audio, and video by utilizing a shared latent space across different modalities.

Language models (LMs) have achieved state-of-the-art results across a wide range of benchmarks in natural language processing, coding, mathematics, and reasoning. They have also demonstrated impressive abilities to generate realistic and creative content, including stories, poems, songs, and jokes. These advancements in LMs have unlocked new opportunities and introduced challenges for AI research and applications, spanning areas such as natural language understanding, natural language generation, multimodal fusion, and knowledge extraction.

However, LMs are not ideally suited to address domain-specific challenges, such as specialized vocabulary, terminology, knowledge, and logic, due to their design for broad and diverse applications. Previous studies highlight these limitations across several domains. In Biomedicine \cite{Gu_2021, Luo_2022}, LMs struggle to capture complex relationships and semantics of biomedical entities and concepts. In Finance \cite{wu2023bloomberggpt}, LMs underperform compared to domain-specific counterparts. In Medicine \cite{singhal2023expertlevelmedicalquestionanswering}, LMs lack alignment with clinical utility. In Security \cite{SecBERT, GoogleCloudAIWorkbench}, LMs fall short in domain-specific security knowledge. Similarly, in Software \cite{10254958}, LMs face challenges in interpreting and making sense of operational logs.

Despite these limitations, the potential of Artificial Intelligence (AI), particularly generative AI, continues to attract significant interest from the security community \cite{securityCopilot, crowdstrikeCharlotteGenerative, SecPalm}. These generative models hold potential as valuable tools for security professionals, serving as copilots to navigate the complexities of security tasks such as identifying phishing emails, crafting detections, or analyzing and summarizing incidents. However, the field of security presents unique challenges that can only be partially addressed by generative models. A notable challenge is the need for security professionals to handle a variety of data types beyond natural language texts, including logs and telemetries. These data are often heterogeneous, noisy, and voluminous, necessitating efficient and scalable processing and analysis methods.

In this paper, we present \name{}, a small language model that is trained with security logs. \name{} is an \emph{encoder-only} model, pretrained on a large corpus of security logs, which capture various events and activities related to security incidents and operations. \name{} aims to demonstrate the feasibility and utility of training a domain-specific language model on security logs at scale, and to provide a versatile and powerful model that can be applied to various security use cases. We evaluate \name{} on both intrinsic and extrinsic tasks, such as log analysis, anomaly detection, log search and incident classification. Our main contributions are:

\begin{itemize}

\item We pretrain a security-specific small language model from scratch on a large and diverse corpus of security logs, which capture various events and activities related to security incidents and operations. We aim to train a versatile and powerful model that can generalize to various type of security use cases. 

\item We evaluate \name{} using both intrinsic and extrinsic measures, using 
both internal and publicly available benchmarks. We also compare \name{} to the other LMs, such as BERT, DeBERTa and OpenAI's embedding model (text-embeddings-ada-002), and show that \name{} outperforms the best results on most of the tasks, and also exhibits some unique and novel capabilities. 

\item We present four real-world use cases for \name{}. Notably, some of these use cases such as incident classification and threat intelligence document retrieval demonstrate that, despite \name{} being primarily trained on logs, it can effectively generalize to other data modalities without specific training on them. This finding suggests that logs could serve as valuable data sources for pretraining language models across domains beyond security.

\item We discuss the limitations and future directions for \name{}, focusing on areas such as data quality and diversity, as well as improvements in robustness and inference speed.

\end{itemize}

The remainder of this paper is structured as follows: Section \ref{sec:related} discusses related work, while Section \ref{sec:system} introduces the overall architecture and design. Section \ref{sec:eval} details the various experiments conducted for testing and evaluation. Section \ref{sec:use_cases} explains multiple real world use cases of \name{} and the corresponding results. Section \ref{sec:discussion} delves into limitations of \name{} and discusses future work, and finally, Section \ref{sec:conclusion} provides the conclusion.

%% file: related.tex
\section{Related Work}
\label{sec:related}

\subsection{Language Models}
Large and small Language Models are a type of deep neural network that contain a substantial number of parameters. These models are trained using vast amounts of data in a self-supervised manner, which enables them to perform a wide range of tasks. These tasks include, but are not limited to, generating text, images, and videos, summarizing content, answering questions, and reasoning.

LMs are typically categorized into three types based on their use of the transformer architecture: encoder-only, decoder-only, or both encoder and decoder models. Encoder-only models, such as BERT \cite{devlin2019bert} and DeBERTa \cite{he2021debertav3, he2021deberta}, are primarily used as representational models. They are capable of converting a given input into numerical values, such as vector representations. This feature makes them particularly useful for finetuning to different tasks.

On the other hand, some of the most popular LMs are decoder-only generative models. Examples of these include Gemini \cite{geminiteam2023gemini, 2024gemini15}, GPT \cite{brown2020language, openai2023gpt4, gpt_4o}, LLama \cite{touvron2023llama, touvron2023llama2}, and PaLM \cite{chowdhery2022palm, anil2023palm}. These models rely on the now well-known Transformer architecture \cite{vaswani2017attention}, with various modifications such as the mixture of experts \cite{JACOBS1991219} to enhance their performance. Beyond the realm of transformers, new architectures are also emerging. One such example is the Mamba model \cite{gu2023mamba}, which represents the continuous evolution and innovation in the field of Language Models.

\begin{figure*}[h!]
    \centering 
    \includegraphics[width=1.0\textwidth]{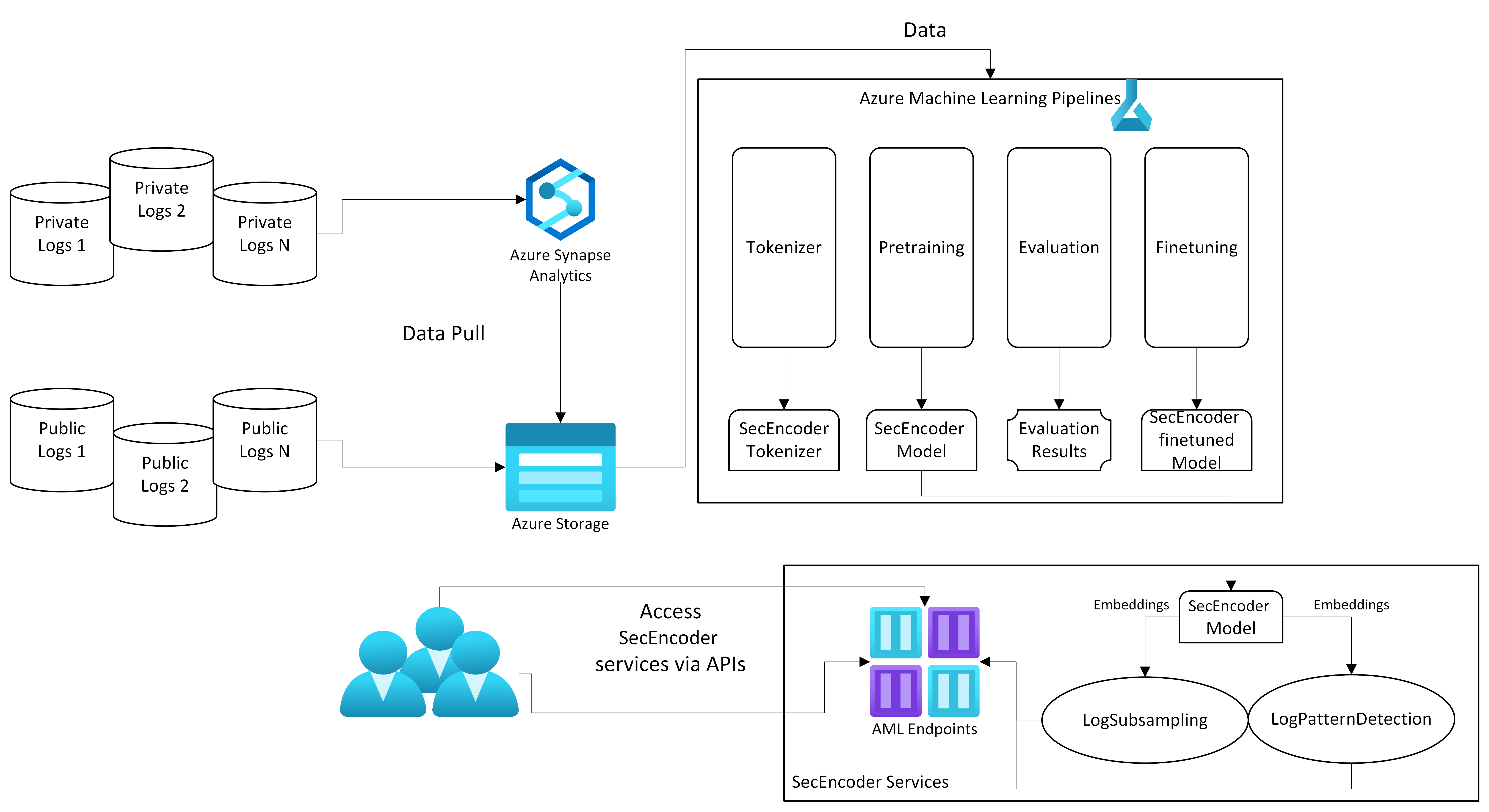}
    \caption{The \name{} architecture encompasses the entire workflow, including data processing, training, evaluation, and deployment.}
    \label{fig:\name{}_arch}
\end{figure*}

\subsection{Domain Adaptation}
While LMs are generally versatile and capable of handling a wide range of tasks without finetuning, domain adaptation may be necessary to achieve optimal performance in specialized fields.

Beyond pretraining a language model from scratch, there are three main techniques for domain adaptation: finetuning, prompt tuning, and in-context learning, which includes few-shot and zero-shot learning. Each approach has its own limitations. Finetuning a pretrained LM for a new domain can result in the loss of important token weights, especially if they are not part of the original vocabulary and the new domain has a different corpus. Additionally, finetuning may lead to catastrophic forgetting \cite{luo2023empirical}, where the model begins to forget tasks it previously performed well.

Prompt tuning \cite{lester-etal-2021-power, li-liang-2021-prefix} also presents challenges, particularly in deciding which parts to update with gradient descent while keeping the rest intact. Lastly, in-context learning, while convenient and straightforward, has its limitations. A pretrained LM may not be able to learn new, unseen knowledge and may still lack domain-specific capabilities.

In the literature, there are several works that propose domain-specific large and small LMs. These include models for biomedicine \cite{Luo_2022, Gu_2021}, molecular biology \cite{molformerxl}, finance \cite{wu2023bloomberggpt}, and security \cite{cybert, bayer2022cysecbert, rahali2021malbert, doc2vec_security, lee2020catbert, guo2021logbert}. 


\subsection{Security specific LMs}
In the study by \cite{bayer2022cysecbert}, the authors introduce a small language model that is pretrained with natural language security texts, as opposed to logs, which is the focus on this paper. Another study, \cite{rahali2021malbert}, demonstrates how to finetune a LM for the purpose of detecting malicious software, highlighting a specific use case. There are also other specific applications of finetuning, which include phishing and threat detection \cite{doc2vec_security, lee2020catbert}, as well as log anomaly detection for specific types of logs such as system logs \cite{guo2021logbert, huang2020hitanomaly, zhang2019robust, ryciak2022anomaly}.

Security is a dynamic domain where attackers are constantly evolving and using different behaviors and vocabularies compared to natural language. \name{} stands out from previous approaches as it focuses on pretraining from scratch at a scale not attempted with other security-specific LMs \cite{bayer2022cysecbert, SecBERT}. The ultimate goal of \name{} is to serve as a general-purpose tool for security logs, rather than being tailored for specific use cases, as targeted in some of the previous work \cite{rahali2021malbert, guo2021logbert, huang2020hitanomaly, zhang2019robust, ryciak2022anomaly, SecBERT}. 

%% file: system.tex
\section{System Design}
\label{sec:system}

Figure \ref{fig:\name{}_arch} illustrates the comprehensive architecture of the \name{} system, detailing the process from initial data extraction to final deployment. In the following section, each component is thoroughly explained, providing in-depth insights into their individual functions and interactions within the overall framework.

\subsection{Data}

\name{} leverages two distinct types of data sources for pretraining: public and private. Public data sources are carefully mined from the Internet, with a focus on those that have permissible licenses to ensure compliance and ethical use. The second type of data utilized for training \name{} comprises private data sources, specifically proprietary data owned by Microsoft (not customer data). This includes security-related data that is instrumental in helping Microsoft operate its business securely and efficiently. Microsoft possesses a vast and rich repository of such telemetry data, as referenced in \cite{microsoft_defender_threat_intelligence}. Appendix \ref{sec:app:training_data} offers a comprehensive overview of the datasets utilized during the pretraining phase of \name{}. In total, we utilized 1 terabyte of data for training \name{}. This data encompasses a diverse set of logs from a wide array of sources, including hosts, devices, systems, and users. It covers various domains such as identity, operating systems (Windows, Linux), cloud, networks, and applications, ensuring a comprehensive and robust training dataset.

\name{} leverages the robust capabilities of Azure Synapse Analytics to efficiently process raw data \cite{microsoft_synapse_analytics}. By harnessing the parallel processing power of Synapse, we can perform comprehensive data preprocessing tasks, including deduplication, for which \name{} employs advanced techniques inspired by the methods described in \cite{brown2020language}. Specifically, our deduplication process encompasses both exact and approximate deduplication strategies, where  exact deduplication identifies and removes duplicate records that are identical in every aspect, while approximate deduplication uses MinHash to detect and eliminate records that are not identical but highly similar. By employing these combined methods, we effectively reduce the dataset size from an initial 1TB to approximately 270GB, which equates to around 77 billion tokens. This reduction ensures a more manageable and efficient dataset for pretraining, especially considering the limited GPU budget.

\subsection{Tokenizer}

The \name{} tokenizer is based on the Byte-Pair Encoding (BPE) algorithm \cite{sennrich2016neuralmachinetranslationrare}. We choose to train the \name{} tokenizer with BPE for several reasons as we outline below:

\begin{itemize} 
\item \textbf{Handling Rare Tokens and Flexibility}: Security logs often contain elements that are challenging for traditional NLP tokenizers, such as IP addresses, timestamps, unique identifiers, or error codes. BPE excels in handling these rare tokens by breaking them down into more manageable subwords, ensuring that even uncommon or unique sequences are effectively tokenized. 
\item \textbf{Vocabulary Size Reduction}: Security logs are typically vast in volume, containing a myriad of unique terms and phrases. By decomposing words into subwords, BPE significantly reduces the overall vocabulary size. This reduction is crucial for efficient processing and storage, as it minimizes the memory footprint and enhances the tokenizer’s performance. 
\item \textbf{Generalization to Unseen Data}: One of the key advantages of BPE is its ability to generalize to unseen data. By learning subwords, the \name{} tokenizer can effectively handle new and previously unseen terms. This capability is particularly important in the context of security logs, where usernames, IP addresses, and other entities frequently change. BPE ensures that the tokenizer remains robust and adaptable to these dynamic elements. 
\end{itemize}

The \name{} tokenizer is being trained on a substantial dataset, comprising 100GB of data that has been equally and randomly sourced from our pretraining dataset. This extensive and diverse dataset ensures that the tokenizer is exposed to a wide range of scenarios and variations, further enhancing its effectiveness and reliability.  The resulting \name{} tokenizer has $29,952$ number of tokens in its vocabulary. Table \ref{tab:tokenizer_logs} illustrates an example tokenization by \name{} and other tokenizers. Notably, Natural Language (NL-based) tokenizers struggle to recognize security specific terms like \emph{TTY}, \emph{USER} and \emph{PWD} often breaking them into incorrect segments.

\begin{table}[h]
\centering
\footnotesize
\begin{tabular}{|p{1.5cm}|p{0.68\linewidth}|}\hline
	\textbf{Tokenizer} & \textbf{Output} \\ \hline
	Original & sudo: root : TTY=unknown ; PWD=/ ; USER=root ; COMMAND=/bin/ip netns identify 4867 \\ \hline
	BERT & su, \#\#do, :, root, :, T, \#\#TY, =, unknown, ;, P, \#\#WD, =, /, ;, US, \#\#ER, =, root, ;, CO, \#\#MM, \#\#AN, \#\#D, =, /, bin, /, i, \#\#p, net, \#\#ns, identify, 48, \#\#6, \#\#7 \\ \hline
	GPT-3 (text-davinci-003) & sudo, :,  root,  :,  T, TY, =, unknown,  ;,  P, WD, =/,  ;,  US, ER, =, root,  ;,  COMM, AND, =/, bin, /, ip,  net, ns,  identify,  48, 67 \\ \hline
	GPT-3.5 \& GPT-4 & sudo, :, root, :, T, TY, =, unknown, ;, P, WD, =/, ;, USER,=root, ;, COMMAND, =/, bin, /ip, net, ns, identify, 486, 7 \\ \hline
	\name{} & sudo, :, root, :, TTY, =, unknown, ;, PWD, =, /, ;, USER, =, root, ;, COMMAND, =, /, bin, /, ip, netns, identify, 4867 \\ \hline
\end{tabular}
\caption{Tokenizer Outputs. Tokens are separated by commas.}
\label{tab:tokenizer_logs}
\end{table}

\subsection{Pretraining}
\name{} is a transformer-based encoder-only architecture. We envision \name{} as a representational model designed to tackle complex cybersecurity challenges and complement universal generative models such as GPT models. \name{} employs the same transformer architecture as DeBERTa-v2 \cite{he2021deberta}. The DeBERTa (Decoding-enhanced BERT with disentangled Attention) architecture enhances BERT and RoBERTa by incorporating a disentangled attention mechanism and an improved mask decoder. This architecture represents each word with two vectors for content and position, computing attention weights separately for these aspects. The enhanced mask decoder integrates absolute positions, improving the model's ability to predict masked tokens and boosting overall performance on NLP tasks. This architecture has demonstrated superior performance across various NLP downstream tasks. \name{} can handle up to $48,000$ tokens as an input, enabling it to process long log lines effectively.

\subsubsection{Pretraining Loss}

\name{} is pretrained using a customized version of masked language modeling (MLM) loss. The MLM objective involves randomly masking tokens in the input sequence and training the model to predict these masked tokens. This task encourages the model to learn contextual representations and understand the relationships between different tokens.

A portion of the logs in our dataset are structured and transformed into a JSONL format, which includes a large number of delimiter tokens such as double quotes. If we randomly mask tokens in these logs without considering the type of tokens, the model will spend a significant amount of time learning to predict the delimiters, which are not as important for understanding the log content.

To address this and improve training efficiency, we propose a customized MLM loss. This customized loss is designed to mask only the content tokens and not the delimiter tokens. This modification ensures that the model focuses on learning the content of the logs and the relationships between the content tokens, rather than the delimiters. This approach has been shown to improve the quality of the learned representations and enhance the model’s performance on downstream tasks.

Here is the formula for the customized MLM loss:
\begin{equation}
	\label{eq:mlm}
	\begin{split}
		x_{\text{masked}} = \{x_i \sim X \setminus X_{delimiter}\}\\
		\mathcal{L}_{\text{MLM}} = -\sum_{i=1}^{N} \log P(x_i | x_{\text{masked}})\\
	\end{split}
\end{equation}

Formula~\ref{eq:mlm} outlines the customized MLM loss, where $x_i$ represents the token at position $i$, $x_{\text{masked}}$ denotes the masked tokens and $x_{\text{masked}}$ is sampled from the input token set that excludes the delimiters. This loss function encourages the model to predict the masked content tokens, enhancing its ability to learn the content of the logs and the relationships between the content tokens. With this customized MLM loss, \name{} can effectively learn to generate high-quality representations of security logs and perform well on a wide range of downstream tasks.

\subsubsection{Pretraining infrastructure}

\name{} leverages Azure Machine Learning (AML) pipelines to streamline and optimize the pretraining process \cite{azure_machine_learning}. By utilizing these pipelines, \name{} ensures a robust and scalable infrastructure capable of handling large-scale data and complex computations. During the pretraining phase, \name{} integrates several advanced tools and frameworks to enhance performance and efficiency as explained next:

\begin{itemize} 

\item \textbf{HuggingFace, Pytorch-Lightning and PyTorch}: This library provides state-of-the-art models and tools for natural language processing. \name{} utilizes HuggingFace and Pytorch-Lightning libraries to easily pretrain \name{} along with PyTorch \cite{huggingface, pytorchlightning, pytorch}.

\item \textbf{DeepSpeed}: This deep learning optimization library is used to scale up training, allowing \name{} to handle larger models and datasets more efficiently while reducing training time and resource consumption \cite{deepspeed}. 

\item \textbf{ONNX Runtime for PyTorch}: By incorporating ONNX Runtime, \name{} accelerates PyTorch models, ensuring faster inference and reduced latency \cite{onnxruntime}.

\end{itemize}

Together, these technologies enable \name{} to achieve faster, more efficient, and scalable machine learning workflows, ultimately leading to better performance and outcomes. We pretrain \name{} using 64xA100 GPUs. Even though we pretrain multiple models, the largest \name{} model, 1.1B billion parameters, is trained on 77B tokens of security logs, and take approximately 4 weeks to complete the pretraining process. 

\subsection{Evaluation}

\name{} utilizes Azure Machine Learning (AML) pipelines to conduct thorough benchmarking and evaluation processes. At the core of this system is a master pipeline, which orchestrates the initiation of multiple benchmarking pipelines in a distributed manner. This distributed approach ensures that \name{} can be evaluated across a wide range of benchmarks efficiently and effectively. The results from these evaluations are meticulously aggregated and compiled into a CSV file, facilitating straightforward analysis, interpretation and comparison over time. Further details on the evaluation can be found in Section \ref{sec:eval}.

\subsection{Finetuning}

\name{} leverages Azure Machine Learning (AML) pipelines for finetuning. The finetuning pipeline is primarily used to generate models based on the pretrained \name{} model for evaluation purposes. The components within these pipelines facilitate the easy development of additional models, enabling the system to address a wide range of use cases efficiently. 


\subsection{Deployment}

Once the \name{} model is pretrained, it is deployed to Azure Machine Learning (AML) as an endpoint to serve various use cases. These use cases, which some of them we discuss in Section \ref{sec:use_cases}, include log subsampling, log pattern detection and incident classification. AML endpoints offer a versatile and scalable solution for deploying machine learning models, allowing for seamless integration with other services and applications. The endpoints support autoscaling, ensuring that the model can handle varying loads efficiently. Additionally, AML provides an intuitive interface and comprehensive tools that make the deployment process straightforward and user friendly.

%% file: evaluation.tex
\section{Experiments}
\label{sec:eval}

\begin{figure}[!htbp]
    \centering
    \includegraphics[width=0.5\textwidth]{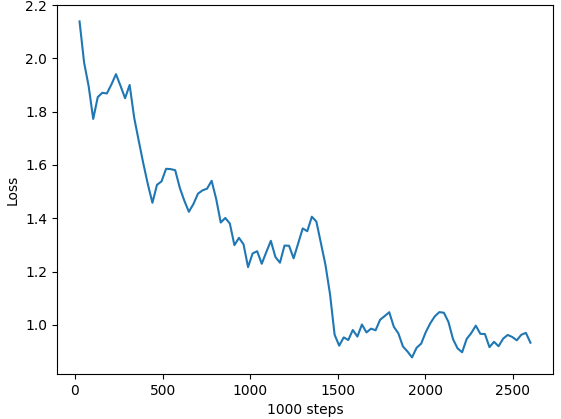}
    \caption{\name{} xxlarge training loss} 
    \label{fig:training_loss}
\end{figure}

\name{} is pretrained in various sizes: base (110M), large (350M), xlarge (700M), and xxlarge (1.1B), following the DeBERTa naming conventions. The base model is trained on 160GB of data (47B tokens), while the large, xlarge, and xxlarge models are each trained on 270GB of data (77B tokens). In general, the training process is smooth, with the loss decreasing steadily over time, despite occasional spikes. Figure \ref{fig:training_loss} shows the training loss trajectory for the xxlarge model. Overall, the training experience is consistent, showing a gradual reduction in loss over time.

\name{} is assessed through various dimensions. Section \ref{subsec:intrinsic} delves into the intrinsic evaluations of \name{}. Section \ref{subsec:extrinsic} examines the extrinsic evaluations of \name{} across different downstream tasks. Lastly, Section \ref{subsec:runtime} discusses the inference time of \name{}.


\subsection{Intrinsic Evaluation}
\label{subsec:intrinsic}

This assessment utilizes two key metrics: perplexity and masked token prediction accuracy. Perplexity, a conventional metric for appraising language models, is used to evaluate how well a model predicts a sample of text. The formula to calculate perplexity is given in Equation \ref{eq:perplexity}, where $(w_1, w_2, \ldots, w_N)$ represents the sequence of words, and $P(w_1, w_2, \ldots, w_N)$ denotes the probability of the sequence of words according to the model. Perplexity gauges the uncertainty of a language model in predicting the next token (such as a word or character) in a sequence --the lower the perplexity, the more capable the model.

\begin{equation}
PP(W) = \exp \left( -\frac{1}{N} \sum_{i=1}^{N} \log P(w_i | w_1, w_2, \ldots, w_{i-1}) \right)
\label{eq:perplexity}
\end{equation}

Accuracy, as an instrinsic metric, on the other hand, quantifies the proportion of masked tokens that are correctly predicted --the higher the accuracy, the better the model performs. The formula to calculate the accuracy is given in Equation \ref{eq:accuracy}.

\begin{equation}
\text{Accuracy} = \frac{\text{Number of Correct Predictions}}{\text{Total Number of Predictions}}
\label{eq:accuracy}
\end{equation}


We conducted intrinsic evaluations on two datasets: the In-distribution test set (IDTS) and the Out-of-distribution test set (ODTS). The total size of IDTS is 2.2GB ($\sim10$M samples), whereas ODTS is 0.03GB ($\sim200$K samples). The key difference between these datasets is that ODTS is sampled from a different distribution than the training dataset, whereas IDTS shares the same distribution as the training dataset -- albeit none is being used while training \name{}. 

\begin{table}[ht]
    \caption{Intrinsic evaluation results}
    \label{tab:intrinsic_experiment}
    \footnotesize
    \centering
    \begin{tabular}{|l|l|l|l|l|l|}
    \hline
                      &            & Base     & Large    & xLarge & xxLarge   \\ \hline
    IDTS           & Perplexity & 2.23 & 2.27    & \textbf{2.16} & 4.79 \\ \hline
                      & Accuracy   & 0.87 & \textbf{0.90} & 0.88 & 0.69 \\ \hline
    ODTS         & Perplexity & 90.85  & \textbf{67.37}  & 95.63 & 421.23 \\ \hline
                      & Accuracy   & 0.33 & \textbf{0.36} & 0.32 & 0.17 \\ \hline
    \end{tabular}
\end{table}
 
Table~\ref{tab:intrinsic_experiment} presents the results. The \name{} model achieved a perplexity of 2.16 and an accuracy of 0.90 on the IDTS, while on the ODTS, it recorded a perplexity of 67.37 and an accuracy of 0.36. These results indicate that \name{} performs well on the IDTS but is less effective on the ODTS, as expected due to different distribution. Among the \name{} models, the Large and xLarge variants outperformed the others, although the differences were marginal, except for xxLarge which we attribute to its much larger parameter size. Additionally, we observed a steady improvement in these metrics throughout the pretraining phase. These intrinsic metrics are primarily used during pretraining to assess the model's progress, alongside other metrics such as loss, to ensure consistent and steady pretraining in each epochs.

\subsection{Extrinsic Evaluation}
\label{subsec:extrinsic}

For extrinsic evaluation, we assess how well \name{} performs on specific downstream tasks. We focus on three primary tasks: log similarity, log search and log anomaly detection. For these tasks, we evaluate the performance of \name{} embeddings. This helps determine whether \name{}'s embeddings represent security logs and can be effectively used for log-related tasks. We also compare \name{}'s embeddings with those from natural language-trained (NL-based) models. To represent a single embedding vector for a given log, we average the embeddings of all tokens.


\subsubsection{Log similarity}

In this task, we evaluate whether the \name{} embeddings can help distinguishing between similar and different logs.
We evaluate on the linux system logs because they do not have a fixed template, and the logs are more diverse.
To test this, logs with the same log template, parsed by the drain parser~\cite{he2017drain}, are randomly grouped together as positive pairs, while logs with different templates are randomly grouped together as negative pairs. We use the drain parser's default settings, tree depth parameter (depth=$4$) and a similarity threshold parameter (st=$0.4$). In total, $1,000$ positive and $1,000$ negative pairs are generated. Difference in cosine similarity between positive and negative pairs using models' embeddings, is used as a metric to evaluate the performance --the greater the difference, the better the performance is.

Table~\ref{tab:log_search_pair_difference} shows the performance of \name{} in comparison to different NL-based models. The row \emph{PMean} represents the mean cosine similarity of positive pairs, \emph{NMean} represents the mean cosine similarity of negative pairs, and \emph{Diff} is the difference between PMean and NMean. The greater the \emph{Diff}, the better the results. For the log similarity task, \name{} shows substantial improvement over NL-based models (0.46 vs. 0.10). Generally, larger \name{} models perform better than smaller ones.

\begin{table*}[!htbp]
    \caption{\name{} performance on Log similarity and Log search}
    \label{tab:log_search_pair_difference}
    \footnotesize
    \centering
    \begin{tabular}{|l|l|l|lll|lll|}
        \hline
         &  &  \multicolumn{4}{c|}{\name{}} & \multicolumn{3}{c|}{NL-based models} \\ \hline
         &  & Base & \multicolumn{1}{l|}{Large} & \multicolumn{1}{l|}{xLarge} & xxLarge & \multicolumn{1}{l|}{BERT} & \multicolumn{1}{l|}{DeBERTa} & OAI-ada-002 \\ \hline
        \multirow{3}{*}{Log similarity} & PMean & 0.94 & \multicolumn{1}{l|}{0.93} & \multicolumn{1}{l|}{0.94} & 0.88 & \multicolumn{1}{l|}{0.98} & \multicolumn{1}{l|}{0.99} & 0.95 \\ \cline{2-9} 
         & NMean & 0.71 & \multicolumn{1}{l|}{0.62} & \multicolumn{1}{l|}{0.68} & 0.42 & \multicolumn{1}{l|}{0.93} & \multicolumn{1}{l|}{0.95} & 0.85 \\ \cline{2-9} 
         & Diff & 0.23 & \multicolumn{1}{l|}{0.31} & \multicolumn{1}{l|}{0.27} & \textbf{0.46} & \multicolumn{1}{l|}{0.06} & \multicolumn{1}{l|}{0.04} & 0.10 \\ \hline
        \multirow{2}{*}{Log search- nlcluster} & MRR & \textbf{0.32} & \multicolumn{1}{l|}{0.31} & \multicolumn{1}{l|}{0.31} & 0.31 & \multicolumn{1}{l|}{0.30} & \multicolumn{1}{l|}{0.31} & \textbf{0.32} \\ \cline{2-9} 
         & MAP & 0.22 & \multicolumn{1}{l|}{0.22} & \multicolumn{1}{l|}{0.22} & 0.22 & \multicolumn{1}{l|}{0.19} & \multicolumn{1}{l|}{0.20} & \textbf{0.24} \\ \hline
        \multirow{2}{*}{Log search - template} & MRR & 0.76 & \multicolumn{1}{l|}{0.76} & \multicolumn{1}{l|}{\textbf{0.76}} & 0.75 & \multicolumn{1}{l|}{0.75} & \multicolumn{1}{l|}{0.74} & 0.75 \\ \cline{2-9} 
         & MAP & 0.69 & \multicolumn{1}{l|}{0.69} & \multicolumn{1}{l|}{0.69} & \textbf{0.69} & \multicolumn{1}{l|}{0.67} & \multicolumn{1}{l|}{0.67} & 0.68 \\ \hline
        \end{tabular}
    \end{table*}

\subsubsection{Log search}

The second extrinsic task used to evaluate \name{} is log search. This task tests whether the model’s embeddings can help find logs with the same template in a reference set, given a query log. Similar to the log similarity experiment, we use linux syslogs and the Drain parser \cite{he2017drain} to parse the logs into templates. The query set contains 4.7k samples, and the reference set contains 5.4k samples. Notably, the query and reference sets do not share templates with \name{}'s training set.

As a variation, instead of using the Drain parser, we adopt a different approach to find similar logs. First, we create natural language (NL) descriptions of the logs using GPT-4 and then cluster the logs into different groups based on these descriptions. We assume that logs within the same clusters are relevant and should be returned in the search. This NL-cluster approach differs from the template-based approach and may better capture semantic similarity.

To evaluate both template and nl-cluster based approaches, Mean reciprocal rank (MRR) and mean average precision (MAP) are used. MRR measures the average of the reciprocal ranks of results for a set of queries as shown in Equation \ref{eq:mrr}, where $|Q|$ is the total number of queries and $\text{rank}_i$ is the rank position of the first relevant log for the (i)-th query. On the other hand, MAP calculates the mean of the average precision scores for a set of queries, as shown in Equation \ref{eq:map} where $|Q|$ is the total number of queries and $\text{AP}(i)$ is the average precision for the $i$-th query. 

\begin{equation}
\text{MRR} = \frac{1}{|Q|} \sum_{i=1}^{|Q|} \frac{1}{\text{rank}_i}
\label{eq:mrr}
\end{equation}

\begin{equation}
\text{MAP} = \frac{1}{|Q|} \sum_{i=1}^{|Q|} \text{AP}(i)
\label{eq:map}
\end{equation}

Table~\ref{tab:log_search_pair_difference} compares the performance of \name{} with various NL-based models. In the template-based search, \name{} shows a slight improvement over NL-based models. However, in the NL-cluster-based search, OpenAI's text-embedding-ada-002 performs slightly better in MAP and similarly in MRR. Although the specifics are unclear, using GPT-4 descriptions for clustering might have given an unfair advantage to OpenAI's embedding model if it is based on the same underlying model. Nonetheless, \name{} still outperforms other NL-based models, such as BERT and DeBERTa, in most metrics.



\subsubsection{Log Anomaly Detection}

In this task, we evaluate \name{} on the anomaly detection task using the datasets, as detailed in Table~\ref{tab:anomaly_maliciousness_datasets}. These datasets include both private and public sources. The M365 dataset is sourced from M365 data center machines, contains red teaming activities labeled as anomalies. The HDFS \cite{10.1145/1629575.1629587, Loghub}, BGL \cite{4273008, Loghub} and Thunderbird \cite{4273008, Loghub} are obtained from LogHub~\cite{Loghub, loghub_github}, featuring various types of logs for anomaly detection. We follow a similar methodology to that described in \cite{9678773} to split the dataset into training and testing sets. For the HDFS dataset, we utilize only a subset of the full dataset. The F5 dataset consists of syslogs from a single F5 device in Azure, with some logs identified as anomalous by experts. Additionally, two more datasets, curated by Microsoft security researchers, include structured Windows logs and unstructured syslogs. Overall, these datasets represent a diverse set, encompassing both structured and unstructured logs, and various types of logs for comprehensive anomaly detection evaluation.


\begin{table*}[t]
    \caption{Anomaly and Maliciousness Detection datasets}
    \label{tab:anomaly_maliciousness_datasets}
    \footnotesize
    \centering
    \begin{tabular}{|l|l|l|l|p{10cm}|}
    \hline
    Type    & Dataset name     & Train samples & Test samples & Details                                                                                                                        \\ \hline
    Private & M365  & $47,644$         & $17,665$        & Windows log where red-teaming behavior are marked as anomalies. \\ \hline
    Private  & F5 syslogs             & 5,318        & 1,330         & Linux syslogs where some logs are marked as anomalies by security experts.                                              \\ \hline
    Public  & HDFS             & $20,021$         & $5,050$         & Hadoop Distributed File System log. Handcrafted rules to identify the anomalies.                                              \\ \hline
    Public  & BGL              & $189,918$        & $47,479$        & Logs from BlueGene/L supercomputer system.                                                                                      \\ \hline
    Public  & Thunderbird      & $399,999$        & $99,999$        & Logs from Thunderbird supercomputer system. The log contains alert and non-alert messages identified by alert category tags. \\ \hline
    Private  & Structured-windows       & $1,440$          & $6,440$        & Windows endpoint logs that exhibit malicious behavior and sequences are labeled as either malicious or non-malicious.                                                        \\ \hline
    Private  & Unstructured-syslogs     & $2,016$          & $8,526$ & Linux syslogs that exhibit malicious behavior and sequences are labeled as either malicious or non-malicious.         \\ \hline
    \end{tabular}
\end{table*}


\textbf{Embedding quality evaluation}
First, we evaluate the quality of the models' embeddings. To achieve this, embeddings are used as features for the log lines. More specifically, given a dataset, each log line is fed into the model as a text, and an embedding is generated as the average of all token embeddings. These embeddings are then used to evaluate \name{} in both supervised and unsupervised settings. 

In the supervised setting, we utilize a simple Long Short-Term Memory (LSTM) network~\cite{schmidhuber1997long} and Graph Neural Networks (GNN)~\cite{scarselli2008graph}. Each sample of data consists of a set of logs. We first process these logs through \name{}, obtaining sets of embeddings. For the LSTM model, we treat each set of logs as a sequence annotated with timestamps and train the LSTM on these sequences of \name{} embeddings. For the supervised GNN model, we represent each set of logs as a graph. We construct the graph based on the Euclidean distance in the embedding space, drawing edges only between nodes that are within the top 1\% of the smallest Euclidean distances. We then train the GNN on these graphs.

For the unsupervised setting, we use an unsupervised GNN as explained in this paper \cite{xu2022contrastive} and Isolation Forest \cite{liu2008isolation}. For the unsupervised setting, each data point contains a set of logs. We first feed the logs into the \name{} and obtain sets of embeddings. Then we train the unsupervised GNN and Isolation Forest algorithms on all the embeddings.

Table~\ref{tab:anomaly_maliciousness} contrasts the performance of various \name{} models with NL-based models like BERT-large, DeBERTa-v3-large, and OpenAI text-embedding-ada-002. Due to the substantial size of the LogHub datasets and limitations on accessing OpenAI's model only via HTTP calls, we obtained results for only a subset of the datasets. 

The \name{} model outperforms NL-based counterparts in both supervised and unsupervised settings. In the unsupervised scenario, \name{} exceeds the NL-based models by an average F-1 score of 20\% (0.36 vs. 0.30). Similarly, in the supervised scenario, \name{} surpasses NL-based models by an average F-1 score of 13\% (0.88 vs. 0.78). These results suggest that \name{} exhibits superior log semantics understanding in both settings compared to NL-based models.

\begin{table*}[!htbp]
    \caption{\name{} F-1 score on embedding quality evaluation for anomaly detection.} 
    \label{tab:anomaly_maliciousness}
    \footnotesize
    \centering
    \begin{tabular}{|l|l|llll|lll|}
        \hline
        Approach                                       & Dataset              & \multicolumn{4}{l|}{\name{}}                                                      & \multicolumn{3}{l|}{NL-based models}                                   \\ \hline
                                                       &                      & \multicolumn{1}{l|}{Base} & \multicolumn{1}{l|}{Large} & \multicolumn{1}{l|}{xLarge} & xxLarge & \multicolumn{1}{l|}{BERT-large} & \multicolumn{1}{l|}{DeBERTa-v3-large} & \multicolumn{1}{l|}{OAI-ada-002} \\ \hline
        \multirow{8}{*}{Supervised LSTM}               & Structured-windows   & \multicolumn{1}{l|}{0.98} & \multicolumn{1}{l|}{0.98}  & \multicolumn{1}{l|}{0.98}   & 0.98    & \multicolumn{1}{l|}{1}    & \multicolumn{1}{l|}{0.98}    & 0.98        \\ \cline{2-9} 
                                                       & Unstructured-syslogs & \multicolumn{1}{l|}{0.96} & \multicolumn{1}{l|}{0.96}  & \multicolumn{1}{l|}{0.96}   & 0.97    & \multicolumn{1}{l|}{0.97} & \multicolumn{1}{l|}{0}       & \multicolumn{1}{l|}{0}           \\ \cline{2-9} 
                                                       & F5 syslogs           & \multicolumn{1}{l|}{0.65} & \multicolumn{1}{l|}{0.84}  & \multicolumn{1}{l|}{0.8}    & 0.91    & \multicolumn{1}{l|}{0.81} & \multicolumn{1}{l|}{0.8}     & 0.87        \\ \cline{2-9} 
                                                       & M365                 & \multicolumn{1}{l|}{0.91} & \multicolumn{1}{l|}{0.93}  & \multicolumn{1}{l|}{0.92}   & 0.79    & \multicolumn{1}{l|}{0.19} & \multicolumn{1}{l|}{0.57}    & 0.3         \\ \cline{2-9} 
                                                       & HDFS                 & \multicolumn{1}{l|}{1}    & \multicolumn{1}{l|}{1}     & \multicolumn{1}{l|}{1}      & 1       & \multicolumn{1}{l|}{1}    & \multicolumn{1}{l|}{1}       & -           \\ \cline{2-9} 
                                                       & BGL                  & \multicolumn{1}{l|}{0.48} & \multicolumn{1}{l|}{0.51}  & \multicolumn{1}{l|}{0.53}   & 0.41    & \multicolumn{1}{l|}{0.55} & \multicolumn{1}{l|}{0.69}    & -           \\ \cline{2-9} 
                                                       & ThunderBird          & \multicolumn{1}{l|}{0.76} & \multicolumn{1}{l|}{0.69}  & \multicolumn{1}{l|}{0.78}   & 0.78    & \multicolumn{1}{l|}{0.73} & \multicolumn{1}{l|}{0.71}    & -           \\ \hline
        \multirow{8}{*}{Supervised GNN}                & Structured-windows   & \multicolumn{1}{l|}{1}    & \multicolumn{1}{l|}{1}     & \multicolumn{1}{l|}{1}      & 1       & \multicolumn{1}{l|}{1}    & \multicolumn{1}{l|}{1}       & 1           \\ \cline{2-9} 
                                                       & Unstructured-syslogs & \multicolumn{1}{l|}{0.96} & \multicolumn{1}{l|}{0.95}  & \multicolumn{1}{l|}{0.97}   & 0.97    & \multicolumn{1}{l|}{0.93} & \multicolumn{1}{l|}{0.93}    & 0.93        \\ \cline{2-9} 
                                                       & F5 syslogs           & \multicolumn{1}{l|}{0.87} & \multicolumn{1}{l|}{0.75}  & \multicolumn{1}{l|}{0.87}   & 0.88    & \multicolumn{1}{l|}{0.75} & \multicolumn{1}{l|}{0.85}    & 0.85        \\ \cline{2-9} 
                                                       & M365                 & \multicolumn{1}{l|}{0.82} & \multicolumn{1}{l|}{0.78}  & \multicolumn{1}{l|}{0.76}   & 0.6     & \multicolumn{1}{l|}{0.78} & \multicolumn{1}{l|}{0.7}     & 0.72        \\ \cline{2-9} 
                                                       & HDFS                 & \multicolumn{1}{l|}{1}    & \multicolumn{1}{l|}{1}     & \multicolumn{1}{l|}{1}      & 1       & \multicolumn{1}{l|}{1}    & \multicolumn{1}{l|}{1}       & -           \\ \cline{2-9} 
                                                       & BGL                  & \multicolumn{1}{l|}{0.96} & \multicolumn{1}{l|}{0.97}  & \multicolumn{1}{l|}{0.97}   & 0.74    & \multicolumn{1}{l|}{0.92} & \multicolumn{1}{l|}{0.56}    & -           \\ \cline{2-9} 
                                                       & ThunderBird          & \multicolumn{1}{l|}{0.75} & \multicolumn{1}{l|}{0.93}  & \multicolumn{1}{l|}{0.77}   & 0.92    & \multicolumn{1}{l|}{0.32} & \multicolumn{1}{l|}{0.87}    & -           \\ \cline{2-9} 
        Supervised                                     & Avg                  & \multicolumn{1}{l|}{0.86} & \multicolumn{1}{l|}{\textbf{0.88}}  & \multicolumn{1}{l|}{\textbf{0.88}}   & 0.85    & \multicolumn{1}{l|}{0.78} & \multicolumn{1}{l|}{0.76}    & 0.71           \\ \hline
        \multirow{8}{*}{Unsupervised GNN}              & Structured-windows   & \multicolumn{1}{l|}{0.82} & \multicolumn{1}{l|}{0.9}   & \multicolumn{1}{l|}{0.88}   & 0.83    & \multicolumn{1}{l|}{0.86} & \multicolumn{1}{l|}{0.65}    & 0.67        \\ \cline{2-9} 
                                                       & Unstructured-syslogs & \multicolumn{1}{l|}{0.56} & \multicolumn{1}{l|}{0.62}  & \multicolumn{1}{l|}{0.65}   & 0.66    & \multicolumn{1}{l|}{0.4}  & \multicolumn{1}{l|}{0.54}    & 0.4         \\ \cline{2-9} 
                                                       & F5 syslogs           & \multicolumn{1}{l|}{0.63} & \multicolumn{1}{l|}{0.62}  & \multicolumn{1}{l|}{0.64}   & 0       & \multicolumn{1}{l|}{0}    & \multicolumn{1}{l|}{0}       & 0.26        \\ \cline{2-9} 
                                                       & M365                 & \multicolumn{1}{l|}{0.14} & \multicolumn{1}{l|}{0.08}  & \multicolumn{1}{l|}{0.14}   & 0.02    & \multicolumn{1}{l|}{0}    & \multicolumn{1}{l|}{0.03}    & 0.02        \\ \cline{2-9} 
                                                       & HDFS                 & \multicolumn{1}{l|}{0.33} & \multicolumn{1}{l|}{0.81}  & \multicolumn{1}{l|}{0.57}   & 0.6     & \multicolumn{1}{l|}{0.03} & \multicolumn{1}{l|}{0.05}    & -           \\ \cline{2-9} 
                                                       & BGL                  & \multicolumn{1}{l|}{0.13} & \multicolumn{1}{l|}{0.14}  & \multicolumn{1}{l|}{0.13}   & 0.26    & \multicolumn{1}{l|}{0.1}  & \multicolumn{1}{l|}{0.14}    & -           \\ \cline{2-9} 
                                                       & ThunderBird          & \multicolumn{1}{l|}{0.16} & \multicolumn{1}{l|}{0.16}  & \multicolumn{1}{l|}{0.2}    & 0.25    & \multicolumn{1}{l|}{0.03} & \multicolumn{1}{l|}{0.14}    & -           \\ \hline
        \multirow{8}{*}{Unsupervised Isolation Forest} & Structured-windows   & \multicolumn{1}{l|}{0.26} & \multicolumn{1}{l|}{0.3}   & \multicolumn{1}{l|}{0.19}   & 0.25    & \multicolumn{1}{l|}{0.45} & \multicolumn{1}{l|}{0.15}    & 0.13        \\ \cline{2-9} 
                                                       & Unstructured-syslogs & \multicolumn{1}{l|}{0.31} & \multicolumn{1}{l|}{0.24}  & \multicolumn{1}{l|}{0.25}   & 0.28    & \multicolumn{1}{l|}{0.43} & \multicolumn{1}{l|}{0.43}    & 0.43        \\ \cline{2-9} 
                                                       & F5 syslogs           & \multicolumn{1}{l|}{0.22} & \multicolumn{1}{l|}{0.16}  & \multicolumn{1}{l|}{0.21}   & 0.19    & \multicolumn{1}{l|}{0.21} & \multicolumn{1}{l|}{0.21}    & 0.3         \\ \cline{2-9} 
                                                       & M365                 & \multicolumn{1}{l|}{0.38} & \multicolumn{1}{l|}{0.4}   & \multicolumn{1}{l|}{0.42}   & 0.39    & \multicolumn{1}{l|}{0.42} & \multicolumn{1}{l|}{0.29}    & 0.17        \\ \cline{2-9} 
                                                       & HDFS                 & \multicolumn{1}{l|}{0.27} & \multicolumn{1}{l|}{0.25}  & \multicolumn{1}{l|}{0.32}   & 0.27    & \multicolumn{1}{l|}{0.27} & \multicolumn{1}{l|}{0.27}    & -           \\ \cline{2-9} 
                                                       & BGL                  & \multicolumn{1}{l|}{0.37} & \multicolumn{1}{l|}{0.25}  & \multicolumn{1}{l|}{0.36}   & 0.38    & \multicolumn{1}{l|}{0.29} & \multicolumn{1}{l|}{0.16}    & -           \\ \cline{2-9} 
                                                       & ThunderBird          & \multicolumn{1}{l|}{0.14} & \multicolumn{1}{l|}{0.12}  & \multicolumn{1}{l|}{0.13}   & 0.14    & \multicolumn{1}{l|}{0.09} & \multicolumn{1}{l|}{0.09}    & -           \\ \cline{2-9} 
        Unsupervised                                   & Avg                  & \multicolumn{1}{l|}{0.34} & \multicolumn{1}{l|}{\textbf{0.36}}  & \multicolumn{1}{l|}{\textbf{0.36}}   & 0.32    & \multicolumn{1}{l|}{0.26} & \multicolumn{1}{l|}{0.23}    & 0.30          \\ \hline
        \end{tabular}
\end{table*}

\begin{table*}[!htbp]
    \caption{Model quality evaluation via finetuning in anomaly detection task.}
    \label{tab:anomaly_finetuning}
    \footnotesize
    \centering
    \begin{tabular}{|l|llll|ll|}
        \hline
                             & \multicolumn{4}{l|}{\name{}}                                                      & \multicolumn{2}{l|}{NL-based models} \\ \hline
                             & \multicolumn{1}{l|}{Base} & \multicolumn{1}{l|}{Large} & \multicolumn{1}{l|}{xLarge} & xxLarge & \multicolumn{1}{l|}{BERT-large}  & DeBERTa-v3 large \\ \hline
        Structured-windows   & \multicolumn{1}{l|}{1}    & \multicolumn{1}{l|}{1}     & \multicolumn{1}{l|}{1}      & 1       & \multicolumn{1}{l|}{1}     & 1       \\ \hline
        Unstructured-syslogs & \multicolumn{1}{l|}{0.34} & \multicolumn{1}{l|}{0.38}  & \multicolumn{1}{l|}{0.39}   & 0.36    & \multicolumn{1}{l|}{0.38}  & 0.18    \\ \hline
        M365          & \multicolumn{1}{l|}{0.86} & \multicolumn{1}{l|}{0.91}  & \multicolumn{1}{l|}{0.81}   & 0.73    & \multicolumn{1}{l|}{0.90}  & 0.18    \\ \hline
        F5 syslogs           & \multicolumn{1}{l|}{0.82} & \multicolumn{1}{l|}{0.84}  & \multicolumn{1}{l|}{0.84}   & 0.84    & \multicolumn{1}{l|}{0.84}  & 0.81    \\ \hline
        HDFS                 & \multicolumn{1}{l|}{1.00} & \multicolumn{1}{l|}{1.00}  & \multicolumn{1}{l|}{1.00}   & 1.00    & \multicolumn{1}{l|}{0.98}  & 0.98    \\ \hline
        BGL                  & \multicolumn{1}{l|}{0.95} & \multicolumn{1}{l|}{0.98}  & \multicolumn{1}{l|}{0.98}   & 0.98    & \multicolumn{1}{l|}{0.29}  & 0.58    \\ \hline
        ThunderBird          & \multicolumn{1}{l|}{0.79} & \multicolumn{1}{l|}{0.69}  & \multicolumn{1}{l|}{0.84}   & 0.82    & \multicolumn{1}{l|}{0.87}  & 0.78    \\ \hline
        Avg                  & \multicolumn{1}{l|}{0.82} & \multicolumn{1}{l|}{0.83}  & \multicolumn{1}{l|}{\textbf{0.84}}   & 0.82    & \multicolumn{1}{l|}{0.75}  & 0.64    \\ \hline
        \end{tabular}
\end{table*}

\textbf{Model quality evaluation via finetuning} 
In this part, we evaluate model quality through finetuning on the anomaly detection task. We freeze the model up to the penultimate layer and only finetune the last layer of the encoder. Each data point contains a set of logs, so we modify the model architecture to first run a forward pass on all the logs in the set, average the token embeddings of each log, concatenate the embeddings of all logs, and then pass the concatenated embeddings to a classification layer. 

Table~\ref{tab:anomaly_finetuning} shows the F-1 score on the test set. The results indicate that \name{} outperforms the best NL-based models on the finetuning task by an average of 12\% (0.84 vs. 0.75) in F-1 score. The results also indicate that \name{} finetuning achieves comparable performance across various \name{} model sizes, with only marginal improvements as model size increases.

\subsection{Runtime performance of \name{}}
\label{subsec:runtime}

We evaluate \name{}'s runtime performance using a single V100 GPU with 16GB of memory. We processed 2,000 logs from six different tables in a demo Microsoft Sentinel environment \cite{microsoft_sentinel}. These logs are semi-structured, containing both categorical data and open text. The average number of tokens per line ranges from 513 to 1052. Table~\ref{tab:runtime} presents the inference speed (tokens per minute) for different model sizes.

\begin{table}[!htbp]
\caption{\name{} Runtime on a single V100 GPU (s)}
\label{tab:runtime}
\footnotesize
\centering
\setlength{\tabcolsep}{4pt}
\begin{tabular}{|l|l|llll|}
\hline
Data source          & \begin{tabular}[c]{@{}l@{}}Avg tokens \\ per line\end{tabular} & \multicolumn{4}{l|}{Inference speed tokens/min}                                                 \\ \hline
                     &                                                                    & \multicolumn{1}{l|}{Base}  & \multicolumn{1}{l|}{Large} & \multicolumn{1}{l|}{xLarge} & xxLarge \\ \hline
Device Network Events  & 1052                                                               & \multicolumn{1}{l|}{35791} & \multicolumn{1}{l|}{18555} & \multicolumn{1}{l|}{9827}   & 7536    \\ \hline
Device Logon Events    & 709                                                                & \multicolumn{1}{l|}{51239} & \multicolumn{1}{l|}{22303} & \multicolumn{1}{l|}{12190}  & 8428    \\ \hline
Email Events          & 635                                                                & \multicolumn{1}{l|}{55794} & \multicolumn{1}{l|}{23289} & \multicolumn{1}{l|}{12854}  & 8232    \\ \hline
Device Registry Events & 762                                                                & \multicolumn{1}{l|}{48500} & \multicolumn{1}{l|}{20198} & \multicolumn{1}{l|}{10992}  & 7159    \\ \hline
Identity Logon Events  & 513                                                                & \multicolumn{1}{l|}{60150} & \multicolumn{1}{l|}{25162} & \multicolumn{1}{l|}{13812}  & 8689    \\ \hline
Sign-in Events             & 786                                                                & \multicolumn{1}{l|}{47492} & \multicolumn{1}{l|}{20769} & \multicolumn{1}{l|}{11190}  & 7732    \\ \hline
\end{tabular}
\end{table}

As illustrated in Table~\ref{tab:runtime}, \name{}'s inference speed decreases as the model size increases. This is expected due to the increased computational complexity associated with larger models. Specifically, for the same model, longer inputs result in slower processing speeds due to the $O(n^2)$ complexity of the transformer architecture. This quadratic complexity means that as the input length (n) increases, the time required for processing grows quadratically.

For real-world deployment, it is essential to balance \name{}'s speed and performance. While larger models may offer better accuracy and more nuanced understanding, they also require more computational resources and time. Therefore, selecting the appropriate model size and optimizing input lengths are crucial steps to ensure efficient and effective deployment.

In summary, our evaluation highlights the trade-offs between model size, input length, and processing speed. These factors must be carefully considered to achieve optimal performance in practical applications.

%% file: use_cases.tex
\section{\name{} Use Cases}
\label{sec:use_cases}

In this section, we delve into the application of \name{} to real-world challenges through two innovative services: LogSubsampling and LogPatternDetection. Additionally, we explore how \name{} effectively generalizes to various modalities, including incidents, alerts and threat intelligence documents.

\subsection{LogSubsampling} 
Logs are a rich source of security data, providing valuable insights into system activities and potential threats. However, security analysts face significant challenges when investigating large quantities of raw logs and transforming them into actionable insights and recommendations. Generative models like GPT can assist in this transformation by turning raw data into meaningful insights and recommendations. However, due to size limitations, LMs cannot effectively process vast amounts of raw logs in their entirety.

While generic methods exist for breaking large datasets into smaller chunks and combining results, these methods often fail to account for the varying importance of different data chunks. This is where LogSubsampling, powered by \name{} embeddings, comes into play. LogSubsampling effectively selects chunks of data that retain the critical information needed for an investigation, while pruning away redundant and uninformative data. This ensures that the most relevant and diverse set of logs is chosen for analysis.

We hypothesize that \name{} embeddings semantically represent security logs and will help in selecting a diverse and informative set of logs for analysis. To prove this we design and experiment with the following details:

\subsubsection{Data}
We evaluate LogSubsampling across various log types, including Windows logs, Identity logs, Device logon events, Sign-in logs, and Alert Info logs. From each dataset, we select 2000 logs and further subsample to 10 logs each.

\subsubsection{Methodology}
For subsampling, we employ a greedy algorithm based on \name{} embeddings. This algorithm selects logs that maximize the variance of the embeddings, ensuring the chosen logs are both diverse and informative. Initially, the algorithm selects the log closest to the center of all logs. It then iteratively selects the log with the maximum minimum distance to the embeddings of the already selected logs. This \emph{Max-Min} strategy ensures the selected logs are maximally diverse and avoids choosing logs that are too similar to those already selected.

\subsubsection{Evaluation}
To assess the performance of \name{} embeddings on LogSubsampling, we compare them against two baseline methods: random sampling and Term Frequency-Inverse Document Frequency with K-Nearest Neighbors (TF-IDF+KNN). For TF-IDF+KNN, we first apply TF-IDF to transform the text data into feature vectors. Then, we use K-Nearest Neighbors (KNN) to cluster the data into \( N \) clusters, where \( N \) represents the target number of subsampled samples. Finally, we select the samples closest to each cluster center, representing the most central examples within each cluster.

We assess the performance of \name{} embeddings using two metrics: entity count and Levenshtein distance. The entity count shows the number of unique entities in the subsampled logs, while the Levenshtein distance measures the edit distance among the subsampled logs. Higher entity counts and Levenshtein distances indicate more diverse and informative subsampled logs. Entity count is measured at the entity level, where each unique value in each column represents a unique entity count. Levenshtein distance is measured at the character level.

As shown in Table~\ref{tab:subsampling}, \name{}'s greedy approach, on average, returns 9\% more entities than random sampling and 12\% more than tf-idf+knn. While entity counts may consider similar entities as different, Levenshtein distance measures similarity at the character level. Additionally, \name{}'s greedy approach outperforms random sampling by 2.2x and tf-idf+knn by 1.7x in Levenshtein distance on average. This demonstrates that \name{} embeddings are effective in selecting diverse and informative logs for analysis.

\begin{table*}[!htbp]
    \caption{\name{} Evaluation on LogSubsampling}
    \label{tab:subsampling}
    \footnotesize
    \centering
    \setlength{\tabcolsep}{3pt}
    \begin{tabular}{|l|l|l|l|l|l|l|l|}
        \hline
                                                                                       &                                                       & Identity      & Sign-in        & \begin{tabular}[c]{@{}l@{}}Alert\\ Info\end{tabular} & \begin{tabular}[c]{@{}l@{}}Device\\ Events\end{tabular} & Windows  & Average    \\ \hline
        \multirow{3}{*}{\begin{tabular}[c]{@{}l@{}}Entity\\ counts\end{tabular}}   
								         & \name{}                                                   			& \textbf{648}  & \textbf{122}  & \textbf{71}                                       & 201                                                    & \textbf{104} & \textbf{229.2}         	\\ \cline{2-8} 
                                                                                       & \begin{tabular}[c]{@{}l@{}}tf-idf\\ +knn\end{tabular} & 544               & 102          	 & 62                                                   & \textbf{274}                                        & 44     		& 205.2      	 \\ \cline{2-8} 
                                                                                       & Random                                                			& 586               & 108           	 & 61                                                   & 251                                                     & 45     		& 210.2   	  \\ \hline
        \multirow{3}{*}{\begin{tabular}[c]{@{}l@{}}Levenstein\\ distance\end{tabular}} 
								         & \name{}                                                   			& \textbf{1635} & \textbf{997} & \textbf{251}                                      & \textbf{3894}                                      & \textbf{142} &  \textbf{1,383.8} 		 \\ \cline{2-8} 
                                                                                       & \begin{tabular}[c]{@{}l@{}}tf-idf\\ +knn\end{tabular} & 1148             & 692           	 & 171                                                  & 1988                                                    & 20     	  &    803.8   	  \\ \cline{2-8} 
                                                                                       & Random                                                			& 1204             & 703               & 142                                                  & 1049                                                    & 16      	  &   622.8    	   \\ \hline
        \end{tabular}
\end{table*}

\subsection{LogPatternDetection}
LogPatternDetection is a tool designed to identify different patterns in log data. It employs an unsupervised implementation of the IsolationForest algorithm, utilizing \name{} embeddings as a featurizer. This section expands on the results discussed in Section \ref{sec:eval} with a real world use case.

The IsolationForest algorithm operates by creating a random subsample of the dataset, which is then used to construct an isolation tree. Each tree is built using \name{} embedding as a feature, with split values selected randomly. The resulting binary tree is analyzed to identify isolated points, which are expected to have fewer splits, indicating potential anomalies. This method is effective for anomaly detection in logs because it does not require labeled data and can handle high-dimensional datasets efficiently.

We hypothesize that \name{} embeddings semantically represent security logs and will help in finding anomalies within a given set. To prove this we design and experiment with the following details:

\subsubsection{Data}

We evaluate LogPatternDetection using a dataset curated by security analysts. This dataset includes logs from the Microsoft Sentinel product, encompassing various log types such as Device Network Events, Device Logon Events, Device Process Events, Device Registry Events, Email Events, Device File Events, Identity Logon Events, and Cloud App Events.
In each table, the security analyst labels 2000 rows of log with the label benign or anomaly. Since LogPatternDetection is an unsupervised method, we use all the data for testing.

\subsubsection{Evaluation}
Accuracy in the top 5 anomalies is used as the evaluation metric. For instance, if 2 out of 5 identified instances are anomalies, the accuracy is 20\%. The performance of \name{} is compared with the baseline method of label encoding combined with the Isolation Forest algorithm. All evaluations are conducted in an unsupervised anomaly detection setting, with no training or fine-tuning involved.

In addition to the baseline method, we also evaluate a hybrid approach that uses label encoding for structured columns and \name{} embeddings for unstructured columns, as defined by the security analysts.

Results indicate that \name{} achieves an accuracy of 0.57, outperforming the baseline accuracy of 0.50 by 14\%. The hybrid model surpasses both approaches, achieving an accuracy of 0.71, which is a 25\% improvement over \name{} alone and a 42\% improvement over the baseline method.

\subsection{Incident classification}

Incidents and alerts are essential components of cybersecurity operations. Incidents refer to probable security breaches or attacks, whereas alerts are notifications of potential security issues that require further investigation. In this context, \name{} is applied to both incidents and alerts, with the dataset primarily consisting of natural language descriptions of these events. Although these data types are not specifically used to train \name{} or are relatively rare, one could argue that they still contain natural language elements similar to those found in some logs.

\subsubsection{Data}
To evaluate \name{}'s performance in incident classification, we use real world incident data. Each incident is associated with a set of alerts, and each alert is linked to a set of evidence. Incidents are labeled based on analyst judgment into three categories: TP (True Positive), FP (False Positive), and BP (Benign Positive). TP indicates real incidents, FP indicates non-real incidents, and BP indicates benign behaviors such as red-teaming exercises. Over a period of 90 days, in total 4914 incidents are used. 85 days of data (4550 incidents) to train the KNN model and the last 5 days of data (364 incidents) is used for testing.

\subsubsection{Methodology} 
\name{} embeddings are used to automatically classify incidents with the goal of reducing the total case volume for analysts. The goal is to develop two models: a high TP (True Positive) model and a high BP/FP (Benign Positive/False Positive) model, both with high precision. The high TP model should rarely misclassify BP/FP as TP, and the high BP/FP model should rarely misclassify TP as BP/FP. 

To achieve this, we finetune K-Nearest-Neighbors (KNN) classifiers on top of \name{} embeddings. These KNN classifiers are trained on the embeddings and labels of the incidents. Specifically, the classifiers are trained to determine whether an incident is TP or not. We optimize the hyperparameters of the KNN classifiers to achieve both a high TP model and a high BP/FP model. As a baseline, we compare \name{} with OpenAI's embedding (text-embedding-ada-002) model. 

\subsubsection{Evaluation} 
The \name{}-based high TP KNN model achieves a precision of 1 and a recall of 0.42, while the high BP/FP model achieves a precision of 1 and a recall of 0.53. Using OpenAI’s embeddings, the TP KNN model achieves a precision of 0.99 and a recall of 0.36, and the BP/FP model achieves a precision of 0.97 and a recall of 0.42. Our real-world evaluations show that the \name{}-based model would reduce the case volume for analysts by 53.6\%, with 0\% mis-suppression of TP cases and 0\% mis-elevation of BP/FP cases. In comparison, the OpenAI's embedding model (text-embedding-ada-002) would reduce the case volume by 37.3\% and mis-suppress 1.2\% of TP cases, resulting in an overall 44\% improvement for the \name{}-based KNN model.

\subsection{Threat Intelligence Document Retrieval}
Retrieval-Augmented Generation (RAG) is a technique used in Large Language Models (LLMs) to enhance their performance in generating relevant and coherent responses. Unlike traditional language models that rely solely on learned patterns from large datasets, RAG combines generative capabilities with context-specific information retrieval. In RAG, the model incorporates a retriever component that fetches relevant information from a knowledge source, such as a document database, based on the input query or context. This retrieved information is then used to augment the generation process of the language model, leveraging external knowledge to provide more accurate and contextually appropriate responses. Embedding plays a crucial role in RAG systems, as it semantically represents the documents to match them to the given query.

This experiment tests \name{}'s embedding for RAG in a different domain: threat intelligence retrieval, where \name{} has not been specifically trained. It compares \name{}'s performance with OpenAI’s embedding model (text-embedding-ada-002).

\subsubsection{Data}
The experiment utilize 2200 articles from Microsoft Defender for Threat Intelligence \cite{microsoft_defender_ti}. A vector database, stores and organizes data in vector embedding form using both title and the content. Vector databases are particularly well-suited for scenarios where understanding semantic similarity between data points is essential. In the vector databases we store the individual threat intelligence articles and the corresponding numeric embeddings, generated by \name{} as well as a baseline text embedding model from OpenAI (text-embedding-ada-002). 

The evaluation dataset contained 519 question/answer pairs related to the threat intelligence landscape, including inquiries about specific events, techniques, and known threat actors. This dataset is initially formed with the help of an LLM and later reviewed by threat intelligence experts. 

\subsubsection{Methodology} 
The methodology involves using Retrieval-Augmented Generation (RAG) to assist a Large Language Model (LLM), specifically GPT-4, in answering a set of 519 questions. For each question, we query a vector database to retrieve the top 10 relevant threat intelligence documents. These documents are then included in the GPT-4 prompt along with the original question. We compare the performance of a vector database built with SecEncoder embeddings against one built with OpenAI embeddings. Additionally, we generate answers directly using the LLM without RAG to assess its performance based on prior knowledge from its training data.

\subsubsection{Evaluation} 
We use BERTScore \cite{zhang2020bertscore}, ROUGE \cite{lin-2004-rouge}, and GPTScore to semantically compare the generated answers to ground-truth answers. BERTScore leverages BERT embeddings and cosine similarity to compute the similarity between source and target text. The ROUGE metric evaluates the overlap between source and target text, with ROUGE-1 focusing on unigrams (individual words) and ROUGE-L on the longest common subsequence. Both BERTScore and ROUGE generate precision, recall, and F-1 scores. Additionally, GPTScore is used to evaluate the relevance of the source and target text, generating a score from 0 to 10, using GPT-4 as an evaluator.

Surprisingly, as shown in Table \ref{tab:ti_rag}, the \name{}-based RAG approach outperforms both the non-RAG and OpenAI-RAG approaches across most metrics. Compared to OpenAI embeddings, \name{} shows a marginal improvement, demonstrating the capacity of SecEncoder embeddings to enhance a large language model’s understanding, even though \name{} is not specifically trained with any threat intelligence data.

\begin{table*}[h!]
 \footnotesize
\centering
\begin{tabular}{|l|c|c|c|c|}
\hline
\textbf{} & BERTScore & Rouge-1 & Rouge-L & GPTScore \\ \hline
No-RAG & (0.55, 0.61, 0.58) & (0.41, 0.23, 0.26) & (0.26, 0.14, 0.16) & 5.10 \\ \hline
Open AI-text-embedding-ada-002& (0.60, 0.65, 0.62) & (0.49, 0.31, \textbf{0.33}) & (0.34, 0.20, 0.22) & \textbf{6.34} \\ \hline
\name{} & (0.60, 0.66, \textbf{0.63}) & (0.49, 0.30, 0.32) & (0.34, 0.21, \textbf{0.23}) & \textbf{6.34} \\ \hline
\end{tabular}
\caption{Comparison of different approaches for threat intelligence document retrieval. The triples in BERT and ROUGE scores are ordered as follows: precision, recall, F-1 score.}
\label{tab:ti_rag}
\end{table*}

%% file: discussion.tex
\section{Discussion, Limitations and Future Work}
\label{sec:discussion}

First, despite our efforts to pretrain \name{} with a diverse set of logs, we believe there is still room for improvement by incorporating an even broader range of logs. In addition to sourcing data from various external sources, we can explore alternative approaches such as synthetic data generation. Generative models excel at creating similar samples with different variations, which can be adjusted by their temperature settings. This capability can be leveraged to generate diverse datasets, especially when access to original sources is limited.

Second, \name{} demonstrates an impressive ability to generalize across different data modalities, including incidents, alerts, and threat intelligence. While there may be some overlap between logs and these other modalities, we propose that logs, like code \cite{aryabumi2024codecodeexploringimpact}, can play a fundamental role in training language models. Just as code instills logical structure and problem-solving skills, security logs provide structured data and patterns (often including code) that enhance a model's ability to detect trends, analyze events, and apply logical reasoning. This capability is essential for proficiency in anomaly detection and delivering valuable security insights. To further enable \name{} to generalize across a broader spectrum of security scenarios, additional experimentation with diverse types of security data will be critical. These efforts will help us assess and address any limitations in \name{}'s capacity to effectively handle varied data types.

Third, one of the objectives of pretraining \name{} is to facilitate easy finetuning. By being compact yet comparable in size to popular small language models, \name{} offers the flexibility to adapt to customers’ specific data needs. However, finetuning can present challenges, such as the risk of catastrophic forgetting, where the model loses information it originally learned. To mitigate this, more experiments are necessary to ensure that \name{} can effectively adapt to diverse datasets while retaining its foundational knowledge. This will help \name{} better serve customers’ varied requirements.

Lastly, there is still room for optimized the deployment of \name{}. We believe that further optimization efforts will significantly enhance both the speed and efficiency of \name{}. Techniques such as model distillation can be employed to streamline the deployment process. Model distillation involves transferring knowledge from a larger, more complex model to a smaller, more efficient one, thereby maintaining performance while reducing computational requirements. By implementing such optimization strategies, we can ensure that \name{} operates more effectively and efficiently in various deployment environments.

%% file: conclusion.tex
\section{Conclusion}
\label{sec:conclusion}

This paper introduces \name{}, a small language model specifically pretrained with security logs. \name{} demonstrates promising capabilities, outperforming traditional NL-based LMs across various benchmarks. Notably, \name{} also exhibits surprising generalization to other data modalities, including incidents, alerts and threat intelligence. Moving forward, our focus will be on enhancing \name{} by incorporating a more diverse dataset and increasing the model's parameter size.

%% file: appendices.tex
\begin{appendices}

\section{Pretraining Dataset}
\label{sec:app:training_data}

\input{pretraining_data.tex}

\end{appendices}

%% file: pretraining_data.tex
\begin{table*}[h!]
\centering
\small
\begin{tabular}{|l|p{5cm}|p{1cm}|p{10cm}|}
\hline
\textbf{Type} & \textbf{Dataset Name} & \textbf{Size (GB)} & \textbf{Details} \\
\hline
Private & Cyber Defense Operation Center diverse network \& windows logs & 26.18  & Diverse set of network and windows logs, coming from controllers and network devices. Diversity is ensured by picking from unique set of devices and event types.  \\
\hline
Private & Cyber Defense Operation Center sequential network \& windows logs & 24.9 & Sequential set of network and windows logs, coming from controllers and network devices.  \\
\hline
Private & Edge + Platform, Devices and Gaming network \& windows logs & 21.51 &  Windows and networks logs coming from devices and hosts. \\
\hline
Private & Event service entries -Azure Resource Manager Logs & 89 & Azure ARM logs capture all write operations (such as PUT, POST, DELETE) performed through the Azure management API endpoints. These logs provide detailed information about the operations, including their start time and whether they succeeded or failed. \\
\hline
Private & Linux Autoconfig Audit Logs & 71 & Data collected by the audit daemon (auditd) on a Linux system.  \\
\hline
Private & Linux Auto Config Audit syscall logs& 97 & Audit collected by the audit daemon (auditd) specifically related to system calls (syscalls). This data is crucial for tracking and analyzing system-level activities. \\
\hline
Private & Linux Infra guest Audit logs& 68 & Audit logs and data collected from Linux guest systems.  \\
\hline
Private & Linux Infra guest Audit syscall logs & 92 & Syscall auditing captures detailed information about system calls made by processes from Linux guest systems. \\
\hline
Private & Linux overlake Audit logs & 63 &  Data collected by the audit daemon (auditd) on a Linux system.\\
\hline
Private & Linux Overlake Audit syscall logs & 100 & Audit collected by the audit daemon (auditd) specifically related to system calls (syscalls). \\
\hline
Private & Windows Autoconfig General Security Events& 43 &  Logs that capture various security-related activities on a Windows system. These events can include successful and failed login attempts, changes to user privileges, and other security-related actions.\\
\hline
Private & Windows Autopilot General Security Events & 18  &  Autopilot is a collection of technologies used to set up and pre-configure new devices, getting them ready for productive use. \\
\hline
Private & Windows Infra Guest General Security Events& 43  & Security events related to guest logons, particularly in the context of network shares and the Server Message Block (SMB) protocol. \\
\hline
Private & Windows Pilotfish General Security Events& 23 & Security events related to the PilotFish integration engine, which is used for configuring and managing data integration solutions \\
\hline
Private & Entra identity logs & 179 & Sign-in logs for users, service principals for authentication and authorization for resource accesses. \\
\hline
Public & Unified Host and Network Data Set \cite{doi:10.1142/9781786345646_001} & 23 & Network and computer (host) events collected from the Los Alamos National Laboratory enterprise network over the course of approximately 90 days. Windows logs are used. \\
\hline
Public & Firewall - Public Security Log Sharing Site \cite{anton_logsharing} & 3.15 & Bundle 5 and 6. Linux Redhat / Fedora, Snort NIDS, iptables firewall and standard Dragon NIDS alert logs, all signatures enabled. Automatic signature update enabled.  \\
\hline
Public & Webapp - Public Security Log Sharing Site \cite{anton_logsharing} & 6.1 & Bundle 1 (httpd, squid and maillog), Bundle 7 (httpd, squid and maillog), Bundle 8 (Standard web proxy log in W3C format (header, tab separated) from BlueCoat web proxy appliance).  \\
\hline
Public & IDS 2018 Intrusion CSVs (CSE-CIC-IDS2018) \cite{solarmainframe_ids_intrusion_2024} & 47.1  & Contains 10 days' log and network pcaps. The network pcaps are used in pretraining. Only Fridays and Thursdays’ pcaps are used in pretraining. Friday and Thursday’s pcaps result is randomly sampled at 50\% to 27G for pretraining. \\
\hline
Public & Web Server Access Logs \cite{dabbas_web_server_access_logs} & 3.3 & Web sever logs contain information on any event that was registered/logged. This contains a lot of insights on website visitors, behavior, crawlers accessing the site, business insights, security issues, and more. \\
\hline
Public & Log files data from online store \cite{chodak2019log} & 1.5 & Web app access logs. Dataset containing one month of log files from an actual and popular Polish online store. This data-set can provide insights to user behavior as well as performance of the online store. \\
\hline
Public & Cloud-based User Entity Behavior Analytics Log Data Set \cite{landauer2022cloud} & 14 & The data set contains log events from real users utilizing a cloud storage suitable for User Entity Behavior Analytics (UEBA). Events include logins, file accesses, link shares, config changes, etc. The data set contains around 50 million events generated by more than 5000 distinct users in more than five years (2017-07-07 to 2022-09-29 or 1910 days) \\
\hline
\end{tabular}
\caption{Training dataset details. The list contains both private (Microsoft) and public data sources. The dataset does not contain Microsoft customers data.}
\label{tab:pretraining_datasets}
\end{table*}